\documentclass[final,1p,times,longtitle,reversenotenum]{elsarticle}
\usepackage{color}
\usepackage{multirow}
\usepackage{braket}
\usepackage{bm}
\usepackage{hyperref}
\usepackage{tabularx}
\usepackage{lineno}
\usepackage{amssymb}
\usepackage{amsmath}
\usepackage{amsthm}

\newcommand{\nuebar}{$\overline{\nu}_{e}$}
\newcommand{\Co}{$^{60}$Co}
\newcommand{\LY}{${\rm ObsLY}$}

\definecolor{darkgreen}{rgb}{0.0, 0.5, 0.0}
\definecolor{orange}{rgb}{1, 0.5, 0.0}

\journal{Nucl. Instr. Meth. A}

\begin{document}


\title{Optimization of the JUNO liquid scintillator composition using a Daya Bay antineutrino detector}

\begin{frontmatter}

\address[1]{Pontificia Universidad Cat\'{o}lica de Chile, Santiago, Chile}
\address[2]{IPHC, Universit\'{e} de Strasbourg, CNRS/IN2P3, F-67037 Strasbourg, France}
\address[3]{Pakistan Institute of Nuclear Science and Technology, Islamabad, Pakistan}
\address[4]{INFN Catania and Dipartimento di Fisica e Astronomia dell Universit\`{a} di Catania, Catania, Italy}
\address[5]{East China University of Science and Technology, Shanghai, China}
\address[6]{Institute of High Energy Physics, Beijing, China}
\address[7]{University of Science and Technology of China, Hefei, China}
\address[8]{Joint Institute for Nuclear Research, Dubna, Russia}
\address[9]{INFN Sezione di Milano and Dipartimento di Fisica dell Universit\`{a} di Milano, Milano, Italy}
\address[10]{Department of Physics, Faculty of Science, Chulalongkorn University, Bangkok, Thailand}
\address[11]{Comenius University Bratislava, Faculty of Mathematics, Physics and Informatics, Bratislava, Slovakia}
\address[12]{University~of~Wisconsin, Madison, Wisconsin 53706, USA}
\address[13]{Department of Physics and Earth Science, University of Ferrara and INFN Sezione di Ferrara, Ferrara, Italy}
\address[14]{Wright~Laboratory and Department~of~Physics, Yale~University, New~Haven, Connecticut 06520, USA} 
\address[15]{INFN Milano Bicocca and University of Milano Bicocca, Milano, Italy}
\address[16]{INFN Sezione di Padova, Padova, Italy}
\address[17]{University of Roma Tre and INFN Sezione Roma Tre, Roma, Italy}
\address[18]{III. Physikalisches Institut B, RWTH Aachen University, Aachen, Germany}
\address[19]{Brookhaven~National~Laboratory, Upton, New York 11973, USA}
\address[20]{IJCLab, Universit\'{e} Paris-Saclay, CNRS/IN2P3, 91405 Orsay, France}
\address[21]{Eberhard Karls Universit\"{a}t T\"{u}bingen, Physikalisches Institut, T\"{u}bingen, Germany}
\address[22]{Department of Physics, National Taiwan University, Taipei}
\address[23]{Universit\'{e} de Bordeaux, CNRS, CENBG-IN2P3, F-33170 Gradignan, France}
\address[24]{National United University, Miao-Li}
\address[25]{Dipartimento di Fisica e Astronomia dell'Universita' di Padova and INFN Sezione di Padova, Padova, Italy}
\address[26]{Centre de Physique des Particules de Marseille, Marseille, France}
\address[27]{Wuhan University, Wuhan, China}
\address[28]{INFN Milano Bicocca and Politecnico of Milano, Milano, Italy}
\address[29]{Shandong University, Jinan, China}
\address[30]{Dongguan University of Technology, Dongguan, China}
\address[31]{Tsinghua University, Beijing, China}
\address[32]{Nanjing University, Nanjing, China}
\address[33]{Institute of Modern Physics, Chinese Academy of Sciences, Lanzhou, China}
\address[34]{Institute of Physics National Chiao-Tung University, Hsinchu}
\address[35]{North China Electric Power University, Beijing, China}
\address[36]{Sun Yat-Sen University, Guangzhou, China}
\address[37]{Forschungszentrum J\"{u}lich GmbH, Nuclear Physics Institute IKP-2, J\"{u}lich, Germany}
\address[38]{Lomonosov Moscow State University, Moscow, Russia}
\address[39]{Universidade Estadual de Londrina, Londrina, Brazil}
\address[40]{Chinese~University~of~Hong~Kong, Hong~Kong, China}
\address[41]{INFN Sezione di Perugia and Dipartimento di Chimica, Biologia e Biotecnologie dell'Universit\`{a} di Perugia, Perugia, Italy}
\address[42]{Universit\'{e} Libre de Bruxelles, Brussels, Belgium}
\address[43]{Siena~College, Loudonville, New York  12211, USA}
\address[44]{Department of Physics and Astronomy, University of California, Irvine, California, USA}
\address[45]{Institute of Physics, Johannes-Gutenberg Universit\"{a}t Mainz, Mainz, Germany}
\address[46]{Suranaree University of Technology, Nakhon Ratchasima, Thailand}
\address[47]{Charles University, Faculty of Mathematics and Physics, Prague, Czech Republic}
\address[48]{Institute for Nuclear Research of the Russian Academy of Sciences, Moscow, Russia}
\address[49]{Department of Physics, University~of~Illinois~at~Urbana-Champaign, Urbana, Illinois 61801, USA}
\address[50]{School of Physics and Microelectronics, Zhengzhou University, Zhengzhou, China}
\address[51]{Lawrence~Berkeley~National~Laboratory, Berkeley, California 94720, USA}
\address[52]{University of Jyvaskyla, Department of Physics, Jyvaskyla, Finland}
\address[53]{Wuyi University, Jiangmen, China}
\address[54]{Guangxi University, Nanning, China}
\address[55]{Harbin Institute of Technology, Harbin, China}
\address[56]{Technische Universit\"{a}t M\"{u}nchen, M\"{u}nchen, Germany}
\address[57]{Department of Physics, Illinois~Institute~of~Technology, Chicago, Illinois  60616, USA}
\address[58]{Institute of Hydrogeology and Environmental Geology, Chinese Academy of Geological Sciences, Shijiazhuang, China}
\address[59]{School of Physics and Astronomy, Shanghai Jiao Tong University, Shanghai, China}
\address[60]{Forschungszentrum J\"{u}lich GmbH, Central Institute of Engineering, Electronics and Analytics - Electronic Systems(ZEA-2), J\"{u}lich, Germany}
\address[61]{Jinan University, Guangzhou, China}
\address[62]{Beijing Normal University, Beijing, China}
\address[63]{Xi'an Jiaotong University, Xi'an, China}
\address[64]{Institute of Experimental Physics, University of Hamburg, Hamburg, Germany}
\address[65]{Beijing Institute of Spacecraft Environment Engineering, Beijing, China}
\address[66]{Department of Physics, University~of~Houston, Houston, Texas 77204, USA}
\address[67]{China Institute of Atomic Energy, Beijing, China}
\address[68]{Center for Neutrino Physics, Virginia~Tech, Blacksburg, Virginia  24061, USA}
\address[69]{Yerevan Physics Institute, Yerevan, Armenia}
\address[70]{Nankai University, Tianjin, China}
\address[71]{Department of Physics, University~of~Cincinnati, Cincinnati, Ohio 45221, USA}
\address[72]{Department~of~Physics, College~of~Science~and~Technology, Temple~University, Philadelphia, Pennsylvania 19122, USA}
\address[73]{Department of Physics, University~of~California, Berkeley, California 94720, USA}
\address[74]{Department of Physics, The~University~of~Hong~Kong, Pokfulam, Hong~Kong, China}
\address[75]{SUBATECH, Universit\'{e} de Nantes,  IMT Atlantique, CNRS-IN2P3, Nantes, France}
\address[76]{College of Electronic Science and Engineering, National University of Defense Technology, Changsha, China}
\address[77]{The Radiochemistry and Nuclear Chemistry Group in University of South China, Hengyang, China}
\address[78]{Tsung-Dao Lee Institute, Shanghai Jiao Tong University, Shanghai, China}
\address[79]{University of Chinese Academy of Sciences, Beijing, China}
\address[80]{Joseph Henry Laboratories, Princeton~University, Princeton, New~Jersey 08544, USA}
\address[81]{Jilin University, Changchun, China}
\address[82]{Xiamen University, Xiamen, China}
\address[83]{School of Physics, Peking University, Beijing, China}
\address[84]{Laboratori Nazionali di Frascati dell'INFN, Roma, Italy}
\address[85]{California~Institute~of~Technology, Pasadena, California 91125, USA}
\address[86]{College~of~William~and~Mary, Williamsburg, Virginia  23187, USA}
\address[87]{Universidad Tecnica Federico Santa Maria, Valparaiso, Chile}
\address[88]{Pontificia Universidade Catolica do Rio de Janeiro, Rio, Brazil}
\address[89]{National Astronomical Research Institute of Thailand, Chiang Mai, Thailand}
\address[90]{China General Nuclear Power Group, Shenzhen, China}
\address[91]{Institute of Electronics and Computer Science, Riga, Latvia}
\address[92]{Iowa~State~University, Ames, Iowa  50011, USA}
\address[93]{Chongqing University, Chongqing, China}
\author[1]{A.~Abusleme\ensuremath{^{\delta,}}}
\author[2]{T.~Adam\ensuremath{^{\delta,}}}
\author[3]{S.~Ahmad\ensuremath{^{\delta,}}}
\author[4]{S.~Aiello\ensuremath{^{\delta,}}}
\author[3]{M.~Akram\ensuremath{^{\delta,}}}
\author[3]{N.~Ali\ensuremath{^{\delta,}}}
\author[5]{F.~P.~An\ensuremath{^{\delta,\mu,}}}
\author[6]{G.~P.~An\ensuremath{^{\delta,}}}
\author[7]{Q.~An\ensuremath{^{\delta,}}}
\author[4]{G.~Andronico\ensuremath{^{\delta,}}}
% ...


\begin{abstract}
\noindent To maximize the light yield of the liquid scintillator~(LS) for the Jiangmen Underground Neutrino Observatory~(JUNO), a 20~t LS sample was produced in a pilot plant at Daya Bay.
The optical properties of the new LS in various compositions were studied by replacing the gadolinium-loaded LS in one antineutrino detector.
The concentrations of the fluor, PPO, and the wavelength shifter, bis-MSB, were increased in 12 steps from 0.5~g/L and $<$0.01~mg/L to 4~g/L and 13~mg/L, respectively.
The numbers of total detected photoelectrons suggest that, with the optically purified solvent, the bis-MSB concentration does not need to be more than 4~mg/L.
To bridge the one order of magnitude in the detector size difference between Daya Bay and JUNO, the Daya Bay data were used to tune the parameters of a newly developed optical model.
Then, the model and tuned parameters were used in the JUNO simulation.
This enabled to determine the optimal composition for the JUNO LS: purified solvent LAB with 2.5~g/L PPO, and 1 to 4~mg/L bis-MSB.
\end{abstract}


\end{frontmatter}

\begin{center}
\ensuremath{^{\mu}}Daya Bay collaboration

\ensuremath{^{\delta}}JUNO collaboration
\end{center}
\section{Introduction}

Liquid scintillator~(LS) detectors readout by photomultiplier tubes~(PMTs) have supported neutrino physics for several decades, from the discovery of neutrinos in the 1950s~\cite{Cowan}, to the precise measurement of the neutrino mixing angle $\theta_{12}$~at KamLAND~\cite{KamLAND2008}, the precise measurement of solar neutrinos at Borexino~\cite{Arpesella:2008mt}, and the observation of the $\theta_{13}$-driven neutrino oscillation at Daya Bay~\cite{DYBPRL2012}.
Given the high light yield, good transparency and relatively low price, this kind of detector is also adopted by the Jiangmen Underground Neutrino Observatory~(JUNO)~\cite{Djurcic:2015vqa,JUNOYB}, which utilizes 20~kt LS with one of the physics goals of determining the neutrino mass ordering.
Since the sensitivity comes from a precise measurement of the fine structure in the oscillated neutrino spectrum, a crucial requirement on the JUNO detector is the excellent energy resolution, $\sim$3\%~at 1~MeV, corresponding to at least 1100~detected photoelectrons~(p.e.) per MeV of deposited energy~\cite{Djurcic:2015vqa}.
The number of detected p.e.~per MeV is referred to as \LY~hereafter.
The higher \LY~is, the better the energy resolution and the physics sensitivity are.
Thus, one of the keys of the JUNO detector development is to maximize \LY.

In recent LS experiments, a widely used solvent is linear alkylbenzene~(LAB), with 2,5-diphenyloxazole~(PPO) as the fluor and p-bis-(o-methylstyryl)-benzene~(bis-MSB) as the wavelength shifter.
The ionization of a charged particle excites the LAB molecules.
A fraction of the excitation energy is transferred to the PPO.
Scintillation photons with peak wavelength of about 360~nm are generated from the de-excitation of PPO molecules.
The total number of photons released by PPO is defined as the initial light yield, which linearly increases with the PPO concentration before reaching 2~g/L.
Above this concentration the increase becomes much less steep.
The wavelengths of most initial photons are shifted to longer than 400~nm by the absorption and re-emission of bis-MSB.
This shift is crucial as the long wavelength avoids spectral self-absorption by the solvent and allows the photons to reach PMTs far away from the energy deposit points.
Eventually, \LY~is a joint effect of the initial light yield, the photon absorption and re-emission during propagation, and the wavelength-dependent PMT quantum efficiency~(QE).
To obtain the maximum \LY, these aspects must be simultaneously optimized.

There have been many studies that independently measured the optical properties of LS, such as Refs.~\cite{LOMBARDI2013133,XiangMeasurement,Cumming:2018hhe} for the initial light yields, Refs.~\cite{Li_2011,Feng:2015tka,Buck:2015jxa} for the transparency.
To completely deal with the competing photon absorption and subsequent re-emission processes of the LS components, a comprehensive optical model was developed and reported in Ref.~\cite{ZHANG2020163860}.
However, the parameters used in the model were obtained from bench-top experiments with a typical detector size of a few centimeters.
Before usage in JUNO, a spherical LS detector with an inner diameter of 35.4~m, the model and its parameters should be validated based on data collected in a larger detector.

This requirement motivated a dedicated LS experiment at Daya Bay.
A LS pilot plant was built by the JUNO collaboration in the underground LS hall of Daya Bay.
One Daya Bay antineutrino detector~(AD)~\cite{DYBDetector, DYBAD} in the Experimental Hall 1~(EH1-AD1) stopped data taking in January 2017.
The 20~t gadolinium-loaded LS~(Gd-LS) in the innermost cylindrical vessel with 3~m in diameter and height was replaced with purified LS produced by the pilot plant.
PPO and bis-MSB concentrations were 0.5~g/L and less than 0.01~mg/L, and then increased in 12 steps to 4~g/L and 13~mg/L, respectively.
The \LY~in the 13 samples was measured at 0.5\% precision level.
The above-mentioned optical model was successfully tuned to the data.
This allowed to identify the optimal scintillator composition for the final JUNO detector.

The structure of this paper is as follows: Sec.~\ref{sec:replacement} describes the Daya Bay AD and the LS replacement experiment. Section~\ref{sec:LY} presents results of the \LY~measurements. Section~\ref{sec:scale} discusses the determination of the JUNO LS composition using the optical model tuned to Daya Bay data.

\section{The LS replacement experiment at Daya Bay}
\label{sec:replacement}

\begin{figure}[!htb]
\begin{centering}
\includegraphics[width=.5\textwidth]{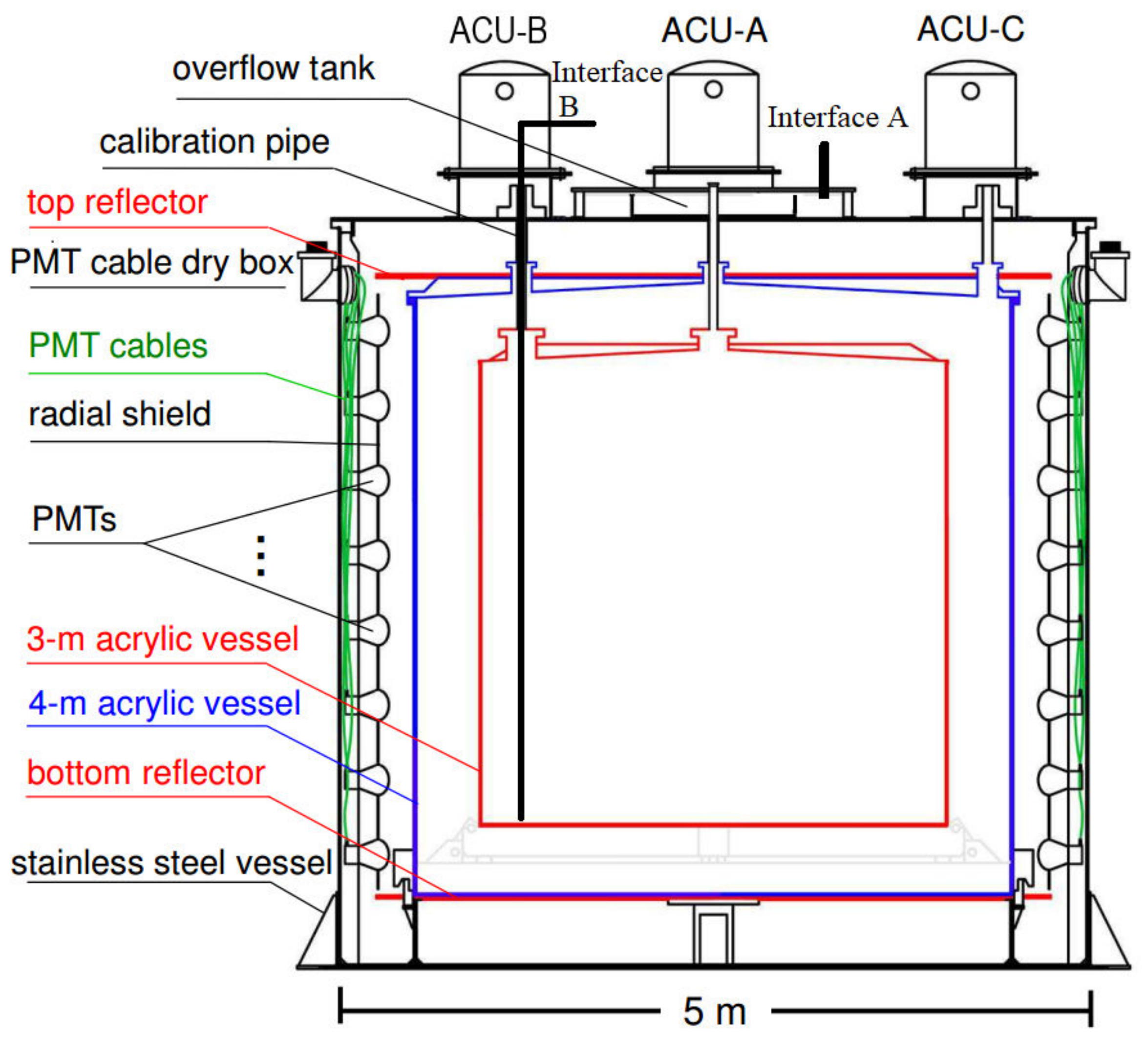}
\caption{\label{fig:AD} Schematic of a Daya Bay antineutrino detector.
The three cylindrical volumes, defined by two acrylic vessels, are filled with Gd-LS, LS, and mineral oil from the innermost to the outermost. Three Automated Calibration Units are installed on top of the detector to calibrate the detector's energy response. Two interfaces are installed on the top of EH1-AD1 for the LS replacement experiment, one through the central overflow tank and the other one using the port of ACU-B.}
\end{centering}
\end{figure}

The Daya Bay reactor neutrino experiment started data taking on December 24, 2011.
With millions of \nuebar~interactions detected in eight identically designed ADs in three underground EHs, many physics results have been produced, including the current world-leading measurements of the neutrino mixing angle $\theta_{13}$ and the squared mass-splitting $|\Delta m^2_{32}|$~\cite{DYBOsc2018}, precise measurements of the reactor \nuebar~flux and spectrum~\cite{DYBSpec2017,DYBEvolution,DYBExtraction}, and stringent limits on the existence of light sterile neutrinos~\cite{DYBSterile2016,Adamson:2020jvo}.
%
%
Three nested cylindrical volumes in each AD are separated by concentric acrylic vessels~(IAV, OAV), as shown in Fig.~\ref{fig:AD}.
The innermost volume is filled with 20~t of gadolinium-loaded LS~(Gd-LS), serving as the primary \nuebar~target.
It is surrounded by $\sim$22~t of non-loaded LS to detect $\gamma$-rays escaping from the target volume.
The outermost volume is filled with mineral oil to shield the LS from natural radioactivity.
A total of 192 8-inch PMTs~(Hamamatsu R-5912) are installed on the steel vessel to detect scintillation photons.
There are three Automated Calibration Units~(ACUs) on the top of each AD to calibrate the energy response along the vertical axes at the detector center~(ACU-A), the edge of the Gd-LS volume~(ACU-B, removed for the replacement experiment), and the LS volume~(ACU-C).
Details of the detector systems are reported in Refs.~\cite{DYBDetector, DYBAD,Dayabay:2014vka}.

\begin{figure*}[tb]
\begin{centering}
\includegraphics[width=1.0\textwidth]{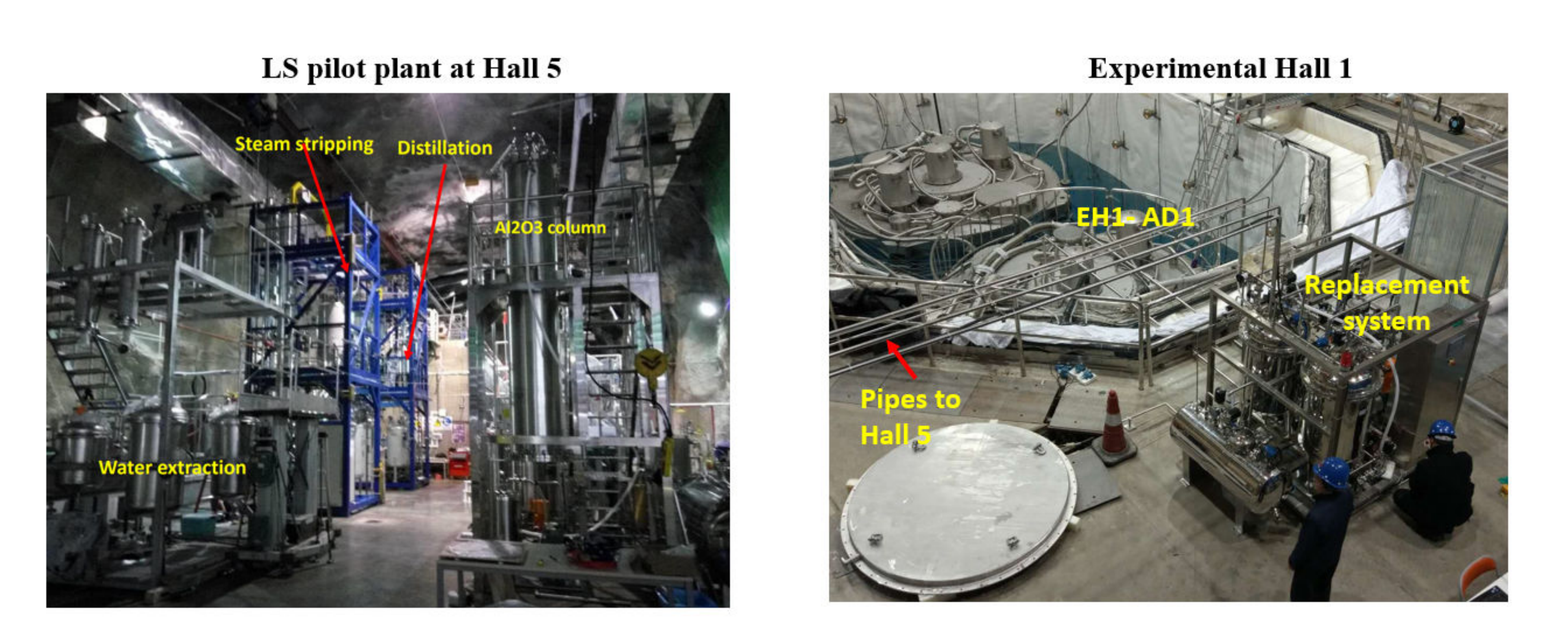}
\caption{\label{fig:setup} Photographs of EH1-AD1, the replacement system in EH1, and the pilot plant in the LS hall~(Hall 5). }
\end{centering}
\end{figure*}


Figure~\ref{fig:setup} shows photographs of the LS experiment, including the pilot plant, the LS replacement system and EH1-AD1.
The pilot plant built in the LS hall consisted of four primary subsystems for purification in sequence: the Al$_2$O$_3$~column, the distillation, the water extraction and the steam stripping.
In addition there were several supporting subsystems for the PPO and bis-MSB dissolution, and the generation of purified water and nitrogen.
The 20~t LAB produced by the Jinling LAB factory was purified by the Al$_2$O$_3$~column for optical transparency and then distillated for radiopurity.
About 11.6~kg PPO produced by the Haiso Technology Co., LTD was dissolved and added to the LAB.
Thus, the initial LS composition was 0.5~g/L PPO without bis-MSB.
Before filling AD1, the mixture went through the water extraction and steam stripping systems for radiopurity.
Details of the distillation and steam stripping systems have been reported in Ref.~\cite{Distillation}.

A replacement system was set up in EH1 to drain the Gd-LS and refill the IAV with new LS.
The system consisted of three pumps, three buffer tanks and stainless steel pipes connecting to the two interfaces on AD1.
The central interface A for injecting liquids was installed via the central overflow tank while the draining was performed via an acrylic tube inserted to the bottom of IAV through the interface B.
The ACU-B was removed for installing the interface B.
In fact, the liquids could be filled or drained from either interfaces.
The system could run in a self-circulation or a full-circulation mode.
In the former mode, liquids were pumped from the IAV to a 300~L buffer tank and then re-inserted into the IAV.
In the latter mode, liquids in the IAV were drained and sent to the facilities in the LS hall for further processing, while newly produced liquids or the re-processed ones were returned to the IAV.

To avoid mixing of the old Gd-LS and the new LS, the Gd-LS was replaced first by purified water that was then replaced by the new LS.
This replacement method was also required to prevent destructive stress on the acrylic vessels.
From February 16 to February 22, the Gd-LS was drained at a rate of about 300~L/h through the central interface while purified water was filled through the side interface.
Then, the new LS with 0.5~g/L PPO was filled through the central interface while water was drained through the side one.
Limited by the position of the central interface A, about 10~L Gd-LS could not be drained out.
This resulted in a residual bis-MSB concentration of less than 0.01~mg/L, confirmed by the light absorption measurement with a UV-vis spectrometer.
In addition, since the tube through the interface B could not touch the IAV bottom, a layer of water with a thickness of about 1~cm was left.
Eventually, the replacement was successfully finished on March 7.
To obtain the radiopurity of LS with 0.5~g/L PPO, the replacement system was shut down after ten days of self-circulation.
The LS radiopurity was measured later in May to wait for the decay of the $^{222}$Rn contamination.

Beginning on May~20 2017, PPO and bis-MSB were added in 12 steps as summarized in Table~\ref{table:steps}.
In each step, the replacement system was working in the full-circulation mode.
The LS was pumped out with a 300~L/h speed and sent to a buffer tank of the water extraction system in the LS hall.
The PPO or bis-MSB was dissolved and slowly added to the buffer tank in 36~hours.
During this time about half of the total LS volume was circulated.
Then, the replacement system ran in the self-circulation mode with a 300~L/h rate for about three days to obtain the uniform fluor distribution in the IAV.
The fluor concentration was measured every 12~hours with a UV-vis spectrometer.
In general, after two days of self-circulation the concentration was stabilized at the target value.
In the following the detector response was calibrated by deploying \Co~calibration sources along ACU-A and ACU-C.
The proceduce generally took about six hours.
\LY~was determined using the data collected with the \Co~source deployed in the detector center.

\begin{table*}[tb]
\begin{centering}
  \small
  \begin{tabular}{c|c|c|c|c|c}
    \hline \hline
	Step & PPO & bis-MSB & Date of calibration & LS temperature&\LY~(p.e./MeV) \\ \hline
	Initial & 0.5~g/L & $<$0.01~mg/L  & April 28, 2017 & 22.6~$^{\rm o}$C &123.7 \\
	1 & 1.0~g/L & $<$0.01~mg/L  & May 28, 2017   & 22.6~$^{\rm o}$C &150.3 \\
	2 & 2.0~g/L & $<$0.01~mg/L  & June 4, 2017  & 22.5~$^{\rm o}$C &167.7 \\
	3 & 2.0~g/L & 0.1~mg/L  & June 9, 2017  & 22.6~$^{\rm o}$C &177.2 \\
	4 & 2.0~g/L & 1.0~mg/L  & June 13, 2017  & 22.6~$^{\rm o}$C &183.2 \\
	5 & 2.0~g/L & 4.0~mg/L  & June 18, 2017  & 22.6~$^{\rm o}$C &184.3 \\
	6 & 2.0~g/L & 7.0~mg/L  & June 23, 2017  & 22.6~$^{\rm o}$C &184.8 \\
	7 & 2.5~g/L & 7.0~mg/L  & June 29, 2017  & 22.6~$^{\rm o}$C &189.5 \\
	8 & 3.0~g/L & 7.0~mg/L  & July 5, 2017  & 22.6~$^{\rm o}$C &191.6 \\
	9 & 3.5~g/L & 7.0~mg/L  & July 11, 2017  & 22.6~$^{\rm o}$C &192.6 \\
	10& 4.0~g/L & 7.0~mg/L  & July 17, 2017  & 22.7~$^{\rm o}$C &192.6 \\
	11& 4.0~g/L & 10.0~mg/L & July 22, 2017 & 22.7~$^{\rm o}$C &193.0 \\
	12& 4.0~g/L & 13.0~mg/L & July 27, 2017 & 22.7~$^{\rm o}$C &193.3 \\
	\hline\hline
  \end{tabular}
  \caption{Summary of the LS experiment. Each LS composition change took 4 to 5 days, including a slow addition of PPO or bis-MSB over 1.5 days followed by at least 3 days of self-circulation. The \LY~was measured to a precision of 0.5\% using a $^{60}$Co calibration source in the detector center. \label{table:steps}}
  \end{centering}
\end{table*}

\section{The light yield measurements}
\label{sec:LY}

The scintillation photons are detected by the 192 PMTs of the AD, operating at an average gain of 1$\times 10^7$.
All the PMTs were working without problems over the three-months period.
In the Daya Bay readout system, after an initial fast amplification, the PMT signal is fed to a pulse shaping circuit consisting of a differential CR and four integrating RC circuits~(CR-(RC)$^4$), and then amplified by a factor of ten.
The integrated value, sampled by a 40-MHz 12-bit ADC, is used as an estimate of the PMT charge output~\cite{DYBFEE}.
Details of the PMT charge calibration are reported in Ref.~\cite{DYBLong}.
The CR-(RC)$^4$ shaping circuit, combined with the time distribution of detected light, introduces a $\sim$10\% nonlinearity in the charge measurement of a single channel, dubbed the electronics nonlinearity.
For the first six PPO/bis-MSB concentrations, this effect was carefully measured based on a full Flash ADC readout system following the method reported in Ref.~\cite{DYBNL}.
For the remaining concentration steps, the Flash ADC system was not working well.
Therefore, the nonlinearity measured at the sixth concentration was used to correct all further measurements, given the time distributions of detected light were found to be stable at these concentrations.

\begin{figure}[!htb]
\begin{centering}
\includegraphics[width=.65\textwidth]{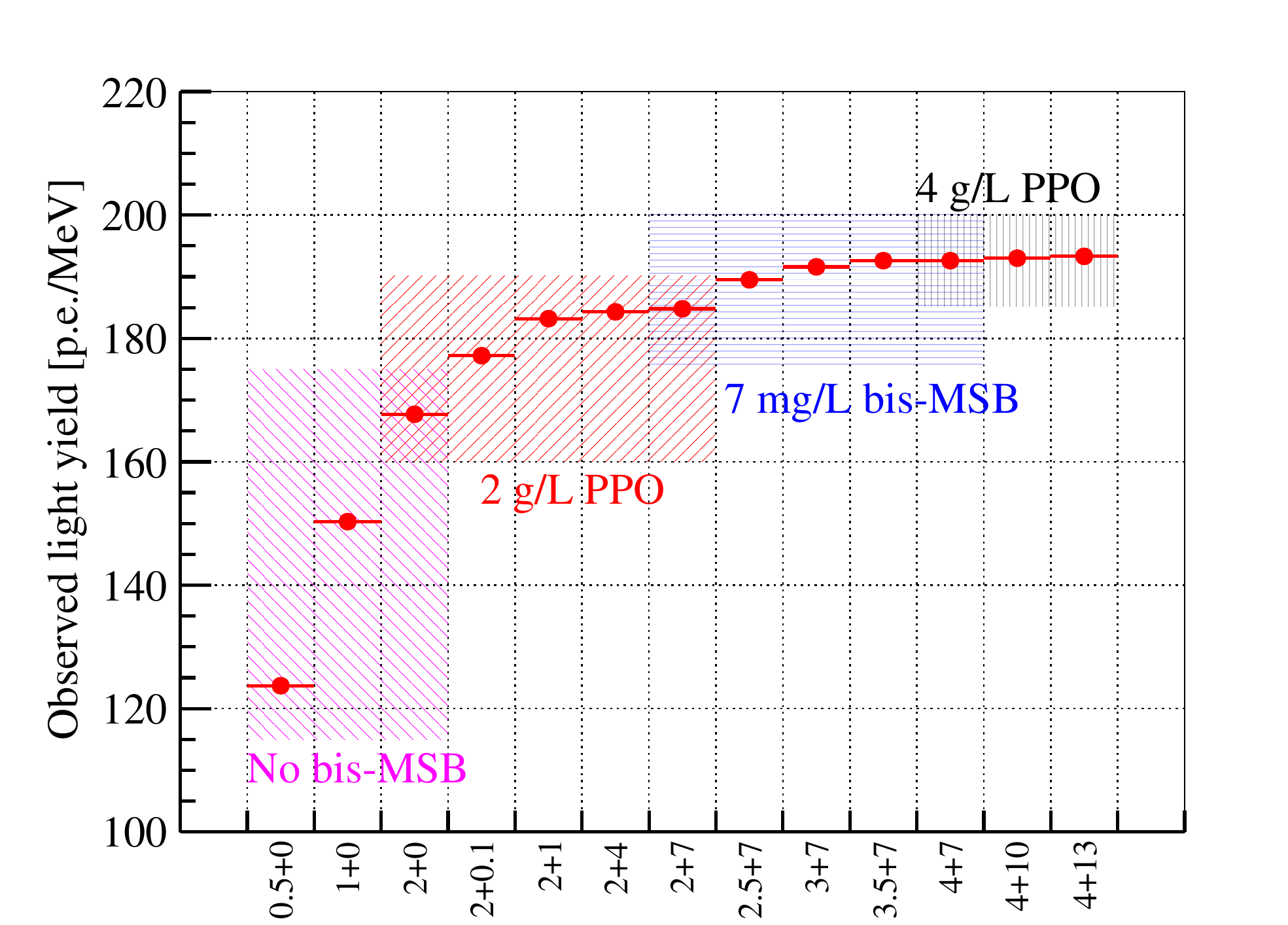}
\caption{\label{fig:LY} Measured light yields \LY~versus PPO and bis-MSB concentrations. The labels of the horizontal axis are the concentrations of PPO~(g/L) plus that of bis-MSB~(mg/L). The vertical error bar~(0.2\%) is statistical only and too small to be visible. The points in each colored box are measured at the same PPO or bis-MSB concentration. }
\end{centering}
\end{figure}

For each concentration, \LY~was determined by measuring the scintillation light originating from the two $\gamma$-rays of $^{60}$Co decays, corresponding to a total deposit energy of 2.505~MeV.
Figure~\ref{fig:LY} and Table~\ref{table:steps} summarize the measured light yields with the 13 LS compositions.
For the first three steps with less than 0.01~mg/L bis-MSB from residual Gd-LS, the light yield increased by more than 40\% with PPO concentrations increasing from 0.5~g/L to 2~g/L.
Adding 1~mg/L bis-MSB further increased the light yield by 10\%.
However, no significant increase was found when further raising the bis-MSB concentration.
This indicates that for the very transparent LAB, scintillation light emitted by PPO would either be absorbed and re-emitted by bis-MSB, or directly reach the PMTs.
Adding more bis-MSB does only shift the fractions of photons from PPO and bis-MSB when reaching the PMTs.
In addition, after adding bis-MSB, the increase of PPO to more than 2.5~g/L yielded no obvious effect on \LY, suggesting the initial light yield had reached the plateau for particles with low energy deposit density, such as $\gamma$'s and $e^\pm$'s.

To compare \LY~of different LS compositions, we have considered the relative uncertainties mainly arising from three sources: the LS temperature variation, the statistical fluctuations in the determination of PMT gains and of the $^{60}$Co peak, and the electronics nonlinearity correction.
The LS temperature was monitored using four sensors and was found to be stable within 0.2~$^{\rm o}$C over the three months, resulting in a less than 0.1\% light yield variation based on the measurements in Ref.~\cite{Dongmei}.
The second term is estimated to be at 0.2\% level according to the fitted energy peak position of the $^{60}$Co source.
The third term is less than 0.5\% for each measurement as discussed in Ref.~\cite{DYBNL}.
Combining the three sources, the uncertainty of each light yield measurement is estimated to be 0.5\%.

\begin{figure}[!htb]
\begin{centering}
\includegraphics[width=.65\textwidth]{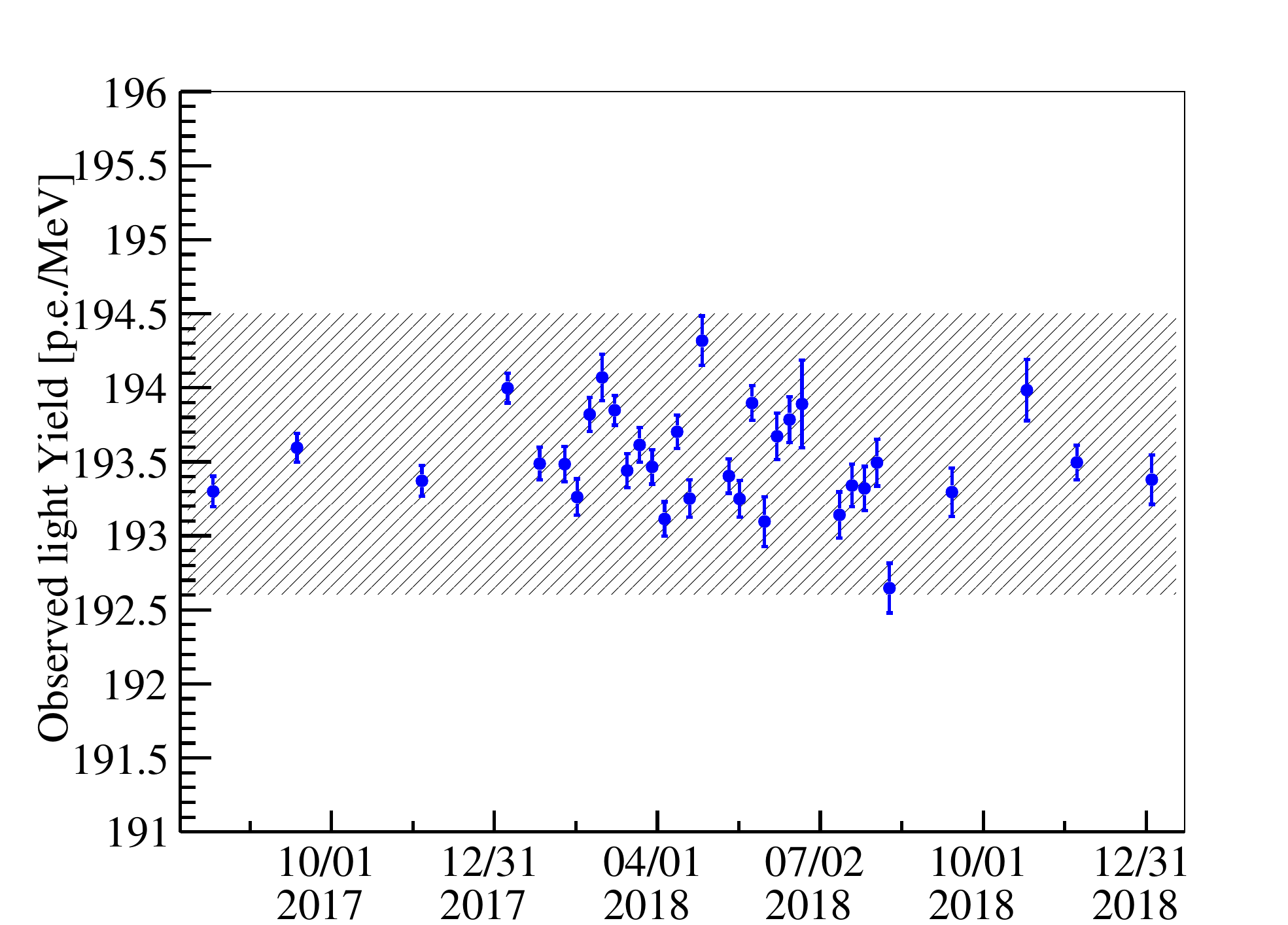}
\caption{\label{fig:LYStability} \LY~of the new LS with 4~g/L PPO and 13~mg/L bis-MSB kept stable within $\pm$0.5\%~(the shaded band) between August 2017 and January 2019. }
\end{centering}
\end{figure}

From August 2017 to January 2019, several rounds of radiopurity studies have been carried out.
In this period, the LS composition was kept at 4~g/L PPO and 13~mg/L bis-MSB.
A stable \LY~was found within $\pm$0.5\% as shown in Fig.~\ref{fig:LYStability}.

\section{Optimization of the JUNO LS composition}
\label{sec:scale}

The \LY~measurements performed at Daya Bay are of great importance to future LS experiments, such as JUNO, for the determination of the LS composition.
However, \LY~of Daya Bay cannot be directly used in JUNO, since \LY~is affected by a few factors, such as the initial light yield, the self-absorption and re-emission effects during propagation, and the PMT QE spectra.
The 20~kt LS of JUNO are contained in a spherical acrylic vessel with an inner diameter of 35.4~m.
Scintillation light is detected by about 18,000 20-inch PMTs, including 5,000 Hamamatsu R-12860 dynode PMTs, and 13,000 NNVT GDG-6201 PMTs with a microchannel plate~(MCP-PMT) instead of a dynode structure.
In Daya Bay the new LS was studied in a cylindrical vessel with 3~m in height and diameter, and Hamamatsu R-5912 8-inch PMTs were used.

To take these differences into account, a newly developed optical model~\cite{ZHANG2020163860} has been implemented in the simulation of both experiments.
The model is designed to thoroughly deal with the competing photon absorption and subsequent re-emission processes of the LS components.
It starts with primary scintillation photons emitted by PPO.
During the propagation, a photon could either be absorbed, or be scattered, or vanish when reaching optical boundaries such as PMTs.
The absorption could happen on any LS component, according to the Beer-Lambert law and the absorption spectrum of each component.
A new photon with longer wavelength may be emitted if the original photon is absorbed by PPO or bis-MSB.
The re-emission probability is defined as the fluorescence quantum efficiency.
In this model, scattering of optical photons happens via the Rayleigh process.
Once a photon is scattered, it changes direction and continues propagation.
The scattering lengths of LAB have been measured in Ref.~\cite{Zhou:2015gwa}.
Thus, key optical parameters in the model consist of emission spectra of PPO and bis-MSB, absorption spectra of LAB, PPO and bis-MSB, and wavelength-dependent fluorescence quantum efficiencies of PPO and bis-MSB.

\begin{figure}[!htb]
\begin{centering}
\includegraphics[width=.7\textwidth]{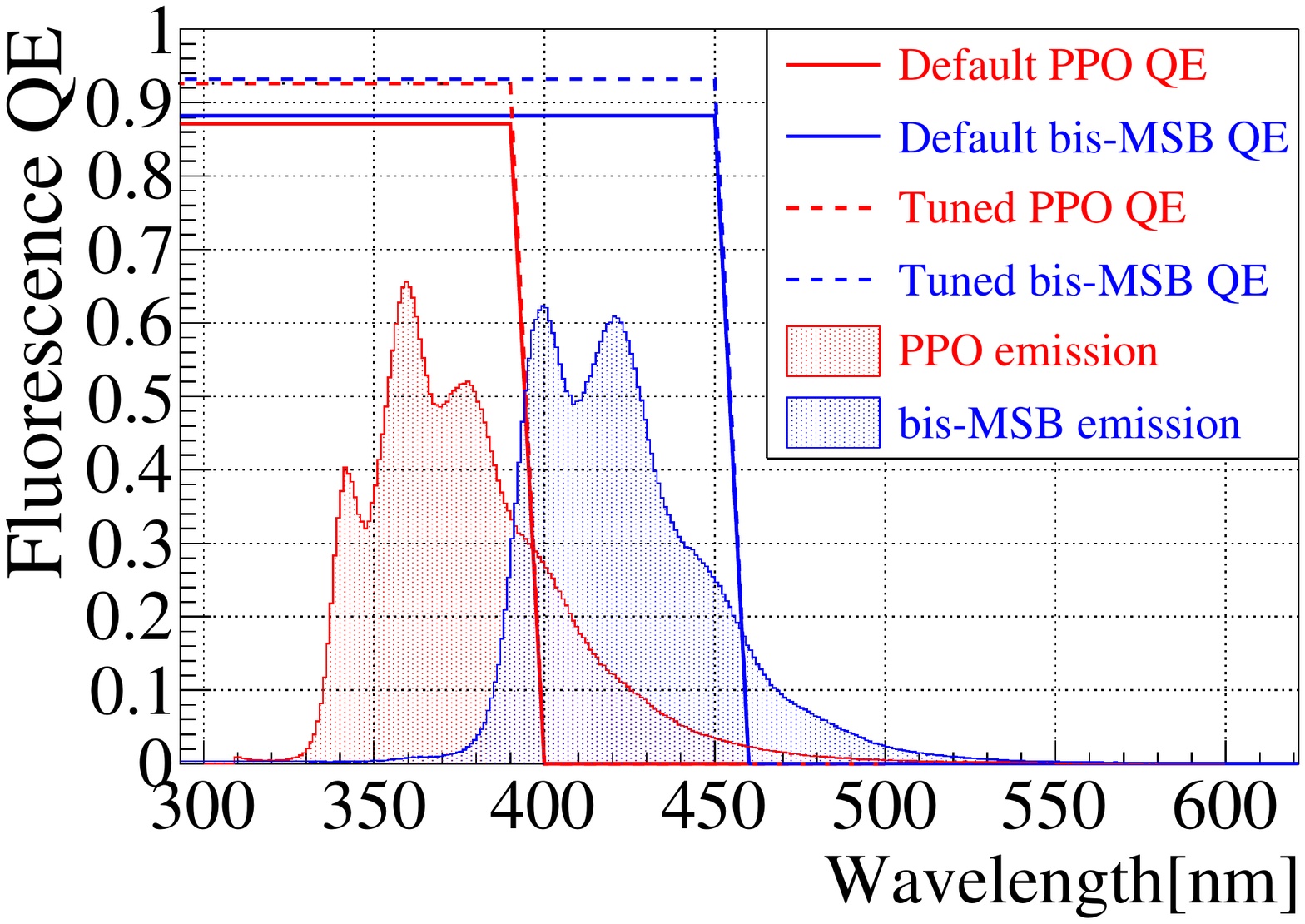}
\caption{\label{fig:OpticalParam} Emission spectra of PPO and bis-MSB in arbitrary units, and their QE efficiencies. The default PPO QE spectra are from the average of three bench-top measurements, while the tuned ones are from the tuning based on Daya Bay data taken in LS experiment in Table~\ref{table:steps}. }
\end{centering}
\end{figure}

\begin{figure}[!htb]
\begin{centering}
\includegraphics[width=.7\textwidth]{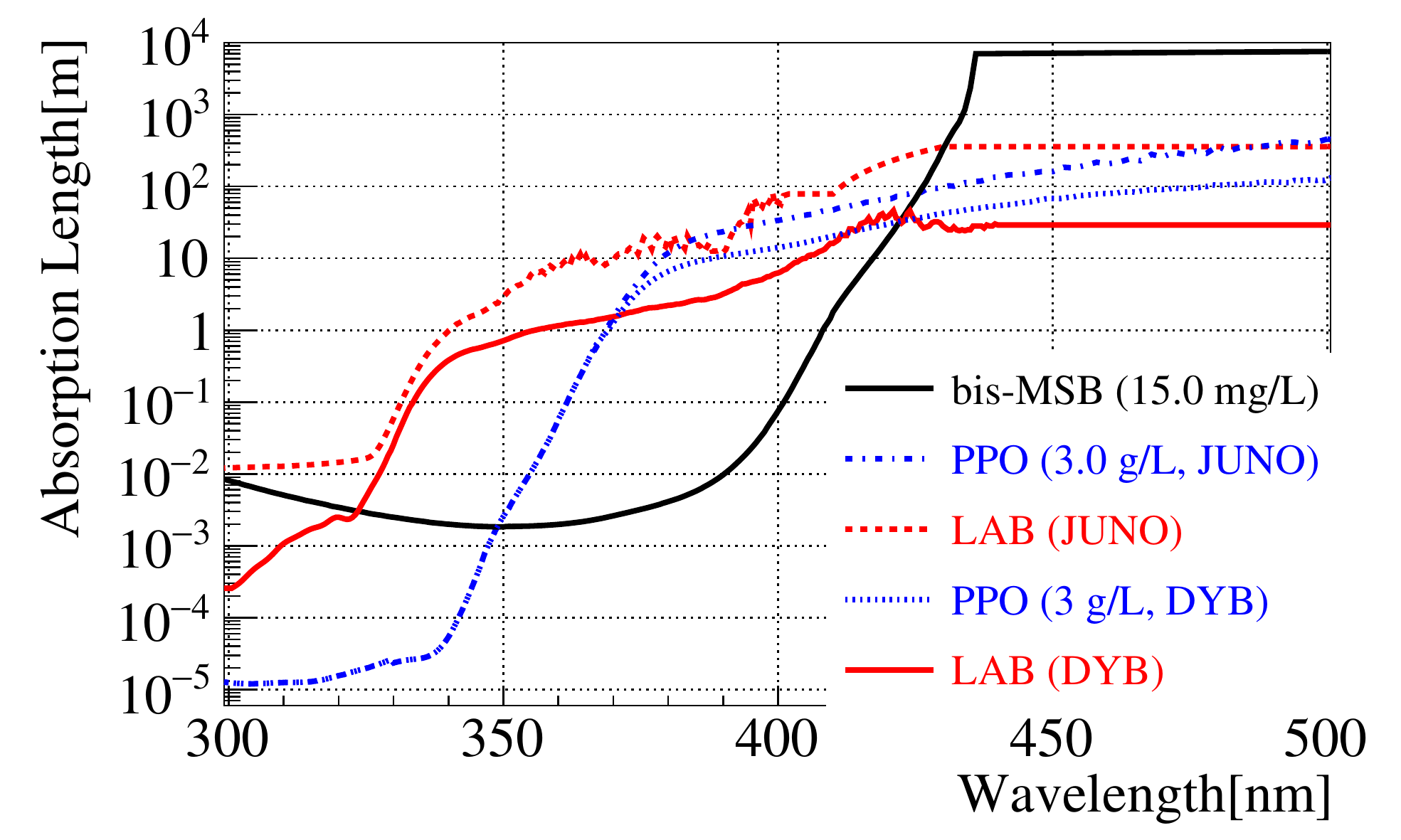}
\caption{\label{fig:LSAbs} Comparison of the absorption spectra of LAB and PPO used in Daya Bay and produced by JUNO pilot plant. The purification significantly improved the optical transparency. }
\end{centering}
\end{figure}

The measurements of the key optical parameters are described below and in Ref.~\cite{ZHANG2020163860}.
The emission spectra of PPO and bis-MSB were well measured with a Fluorolog Tau-3 spectrometer as shown in Fig.~\ref{fig:OpticalParam}.
The absorption spectrum of each LS component used in Daya Bay and the pilot plant was measured using a Shimadzu UV2550 UV-vis spectrometer and quartz cuvettes with different light paths up to 10~cm.
To overcome the large uncertainty due to limited cuvette sizes and the long absorption length at photon wavelength of 430~nm, a 1-m long tube was used.
Comparison of absorption spectra among the Daya Bay original liquids and the newly produced ones is shown in Fig.~\ref{fig:LSAbs}.
The purification significantly improved the transparencies of LAB and PPO.
The general method to measure fluorescence QE was using the combination of a fluorescence spectrometer and a UV-Vis spectrometer.
An average QE spectrum from several measurements~\cite{Li_2011,Feng:2015tka,Buck:2015jxa} was adopted in the simulation, shown as the default QE spectra in Fig.~\ref{fig:OpticalParam}.
Due to intrinsic difficulties of several corrections in this method, the measured efficiencies had relatively large uncertainties, typically 5\%.

\begin{figure}[!htb]
\begin{centering}
\includegraphics[width=.6\textwidth]{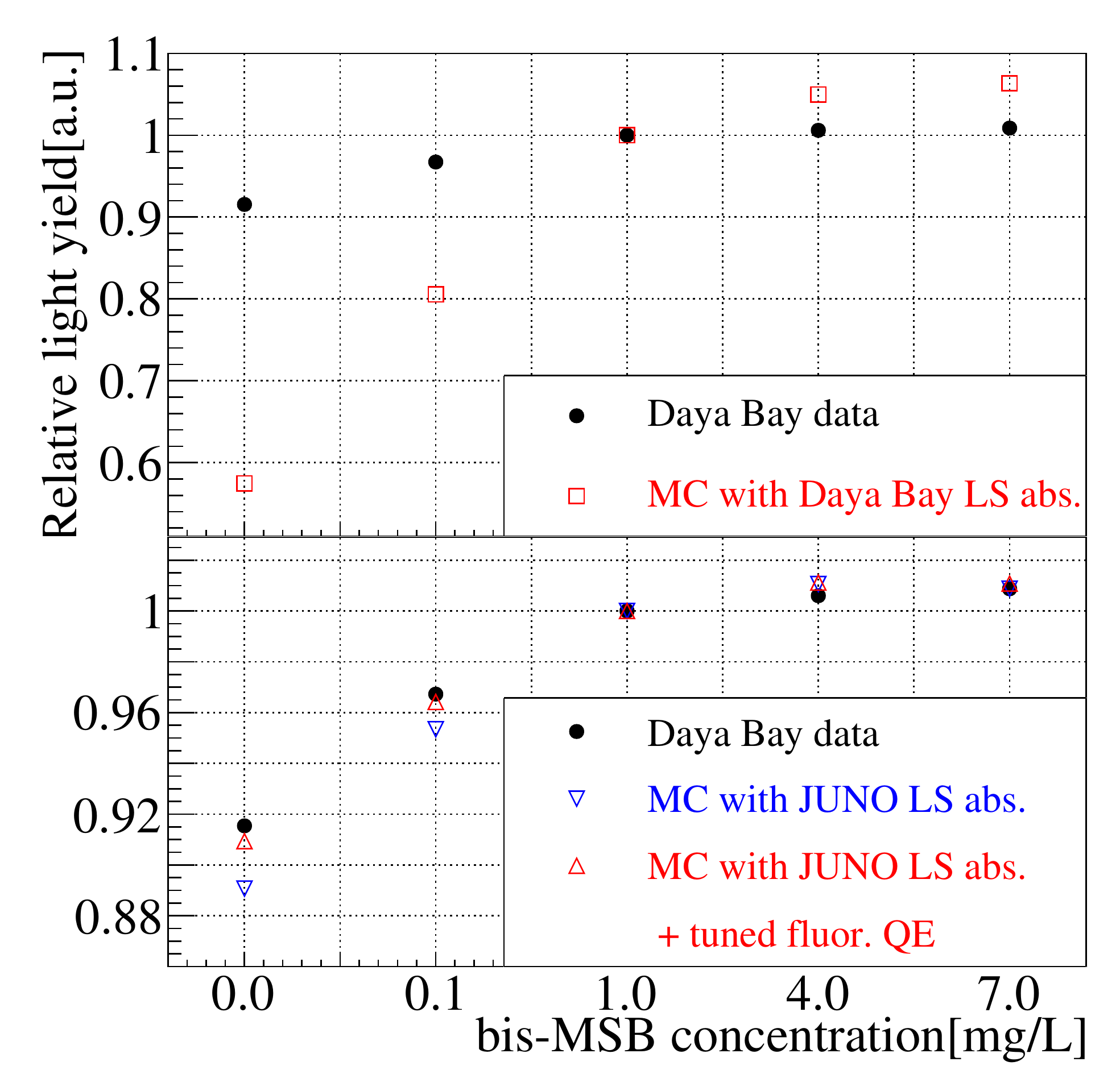}
\caption{\label{fig:DYBTuning} Comparison of the measured and simulated light yields with 2~g/L PPO and various bis-MSB concentrations. Each group is normalized at 1~mg/L bis-MSB. After increasing the PPO fluorescence quantum efficiencies, compared to the data~(black dots), the agreement is improved from 2\%~(blue inverted triangles) to 1\%~(red triangles) at low bis-MSB concentrations. }
\end{centering}
\end{figure}

The optical model has been implemented in the Geant4~\cite{Geant4} based Daya Bay simulation.
The residual 1-cm water at the bottom of IAV has been included in the simulation by assuming a perfect surface between the water and the LS.
The bis-MSB of less than 0.01~mg/L from the residual Gd-LS is also included.
The simulated light yields with respect to bis-MSB concentrations are compared with the experimental data, as shown in Fig.~\ref{fig:DYBTuning}.
If the absorption spectra of Daya Bay liquids are used, a much steeper \LY~increase is found, because LAB, which does not re-emit a new photon after the absorption, strongly competes with bis-MSB in the wavelength range of 350 to 400~nm.
More bis-MSB leads to more photons shifting to the wavelength range above 420~nm, in which the liquids are much more transparent.
Thus, the \LY~monotonically increases with the bis-MSB concentration.
Once the absorption spectra of JUNO liquids are employed, the importance of bis-MSB is significantly reduced, and most of the photons emitted by PPO could reach the Daya Bay LS contained in the outer acrylic vessel before absorption by LAB.
However, at bis-MSB concentrations of smaller than 1~mg/L, the bis-MSB plays a less important role in the data compared to the simulation.
Varying the absorption spectra of each component at wavelengths longer than 420~nm does not reduce the discrepancy.
Changing the height of the bis-MSB QE spectrum, and shifting the cutoff position of the PPO and bis-MSB QE spectra have minor impacts on the discrepancy.
Eventually, the PPO fluorescence QE is increased by 5\% as shown in Fig.~\ref{fig:OpticalParam}.
Once a photon is absorbed by PPO, the probability of re-emitting a new photon with longer wavelength is closer to 1.
In this way, the discrepancy between simulation and data is improved from about 2\% to better than 1\%.
%
%
%

The first application of the tuned model is to extract the initial light yields with different PPO concentrations.
As mentioned before, \LY~is a joint effect of the initial light yield, the absorption and re-emission, and the PMT response.
A set of simulation is performed for the Daya Bay AD with LS compositions with different PPO concentrations and 7~mg/L bis-MSB.
The same initial light yield is used in the simulation to solely study the self-absorption effect of PPO.
The simulation results are listed in Table~\ref{table:simPPO}.
Each 0.5~g/L PPO increase leads to about 0.5\% loss of \LY~in the simulation.
Thus, the initial light yields are obtained by dividing the Daya Bay measured \LY~with the simulated ones, and will be used in the determination of the JUNO LS composition.

\begin{table}[h]
\begin{centering}

  \small
  \begin{tabular}{cccc}
	\hline
    \multirow{2}{*}{PPO concentration} & \multicolumn{2}{c}{\LY} & \multirow{2}{*}{Initial light yield} \\
	\cline{2-3}
	&Measured & Simulated & \\ \hline
	2.0~g/L &1 & 1 & 1 \\
	2.5~g/L &1.025 & 0.994 & 1.031 \\
	3.0~g/L &1.037 & 0.991 & 1.046 \\
	3.5~g/L &1.042 & 0.986 & 1.057 \\
	4.0~g/L &1.042 & 0.982 & 1.061 \\
	\hline
  \end{tabular}
  \caption{ \label{table:simPPO}Relative \LY~with respect to PPO concentrations in the data and the simulation, normalized at 2~g/L PPO. In the simulation the initial light yield is fixed to solely study the PPO self-absorption effects. The initial light yields are extracted by dividing the measured values with the simulated ones. }
  \end{centering}
\end{table}

\begin{figure}[!htb]
\begin{centering}
\includegraphics[width=.6\textwidth]{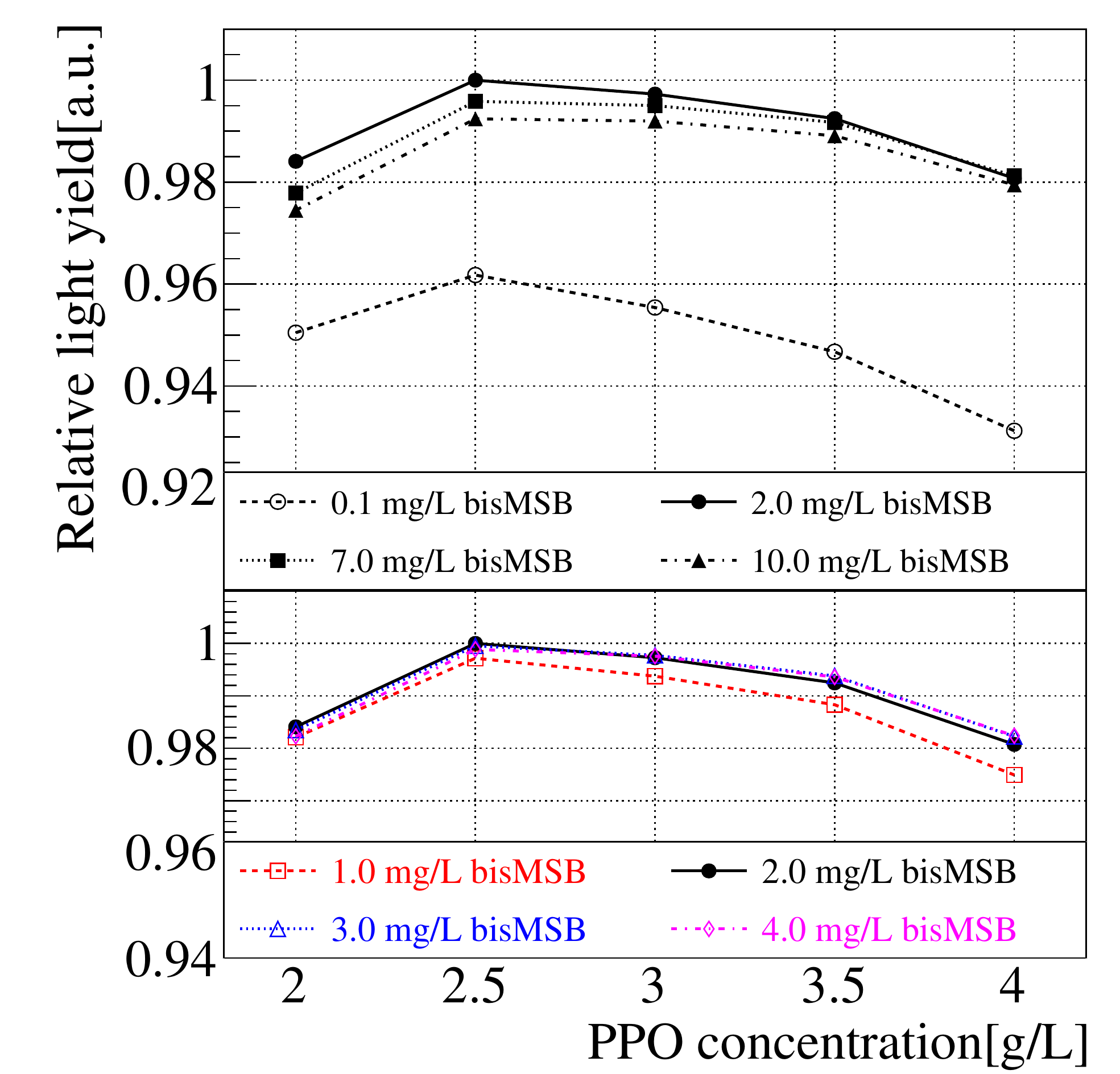}
\caption{\label{fig:JUNOPre} The simulated \LY~with respect to PPO and bis-MSB concentrations in the JUNO detector.}
\end{centering}
\end{figure}

The optical model, the measured absorption spectra, the tuned fluorescence quantum efficiencies, and the extracted initial light yields have been employed in the JUNO simulation.
The simulated \LY~with respect to PPO and bis-MSB concentrations is shown in Fig.~\ref{fig:JUNOPre}.
The normalization point is chosen as 2.5~g/L PPO and 2~mg/L bis-MSB, at which the largest light yield is found.
Although the \LY~monotonically increases with PPO concentrations at Daya Bay, 2.5~g/L PPO is preferred at JUNO due to the non-negligible self-absorption in the larger detector.
The optimal bis-MSB concentration could be in the range of 1~mg/L to 4~mg/L, since the \LY~difference is less than 1\% in this range.

\section{Summary}

A precise measurement of \LY~in various LS compositions has been performed in a Daya Bay AD, by replacing the Gd-LS with purified LS produced in a pilot plant in the underground LS hall.
For $\gamma$'s and $e^\pm$'s, \LY~reaches a plateau for PPO concentrations larger than 2.5~g/L.
In addition, if the solvent is optically purified, the \LY~increase with respect to bis-MSB concentrations is negligible for concentrations larger than 4~mg/L.
A novel optical model has been employed to describe the complicated optical process in the liquids.
The predicted \LY~in different bis-MSB concentrations agrees with these measurements within 1\%.
The initial light yields in various PPO concentrations are extracted by subtracting the PPO self-absorption effect predicted by the optical model.
To find the optimal LS composition of JUNO, the model and the tuned optical parameters are used in the JUNO simulation.
The results suggest that in the JUNO-scale detector, either a PPO concentration larger than 2.5~g/L, or a bis-MSB concentration bigger than 4~mg/L, would reduce the \LY.
The JUNO LS composition is optimized to be the purified LAB with 2.5~g/L PPO and (1-4)~mg/L bis-MSB.
The optimization method can also be used in other future LS experiments.

\section{Acknowledgements}

\label{Acknowledgements}
We are grateful for the ongoing cooperation from the China General Nuclear Power Group
and China Light and Power Company.

Daya Bay is supported in part by
the Ministry of Science and Technology of China,
the U.S. Department of Energy,
the Chinese Academy of Sciences,
the CAS Center for Excellence in Particle Physics,
the National Natural Science Foundation of China,
the Guangdong provincial government,
the Shenzhen municipal government,
the China General Nuclear Power Group,
the Research Grants Council of the Hong Kong Special Administrative Region of China,
the MOE in Taiwan,
the U.S. National Science Foundation,
the Ministry of Education, Youth, and Sports of the Czech Republic,
the Charles University Research Centre UNCE,
the Joint Institute of Nuclear Research in Dubna, Russia,
the National Commission of Scientific and Technological Research of Chile,
We acknowledge Yellow River Engineering Consulting Co., Ltd., and China Railway 15th Bureau Group Co., Ltd., for building the underground laboratory.

JUNO is supported by
the Chinese Academy of Sciences,
the National Key R\&D Program of China,
the CAS Center for Excellence in Particle Physics,
the Joint Large-Scale Scientific Facility Funds of the NSFC and CAS,
Wuyi University,
and the Tsung-Dao Lee Institute of Shanghai Jiao Tong University in China,
the Institut National de Physique Nucl\'eaire et de Physique de Particules (IN2P3) in France,
the Istituto Nazionale di Fisica Nucleare (INFN) in Italy,
the Fond de la Recherche Scientifique (F.R.S-FNRS) and FWO under the ``Excellence of Science ¨C EOS¡± in Belgium,
the Conselho Nacional de Desenvolvimento Cient\'ifico e Tecnol\`ogico in Brazil,
the Agencia Nacional de Investigaci\'on y Desarrollo in Chile,
the Charles University Research Centre and the Ministry of Education, Youth, and Sports in Czech Republic,
the Deutsche Forschungsgemeinschaft (DFG), the Helmholtz Association, and the Cluster of Excellence PRISMA+ in Germany,
the Joint Institute of Nuclear Research (JINR), Lomonosov Moscow State University, and Russian Foundation for Basic Research (RFBR) in Russia,
the MOST and MOE in Taiwan,
the Chulalongkorn University and Suranaree University of Technology in Thailand,
and the University of California at Irvine in USA.

\bibliographystyle{elsarticle-num}
\bibliography{LightYield}

\end{document}